\documentclass[12pt]{article}
\usepackage{cite}
\newtheorem{theorem}{Theorem}
\newtheorem{proposition}{Proposition}
\newtheorem{corollary}{Corollary}

\def\X#1{\mbox{\raisebox{1.2pt}{$\stackrel{#1}{X}$}}}

\begin{document}
\title{\bf Group foliation and non-invariant solutions of the heavenly
equation}
\author{\bf L. Martina$^*$, M.\,B. Sheftel$^\dag$ and P. Winternitz$^\ddag$}
\date{}
\maketitle \noindent ($^*$ Dipartimento di Fisica
dell'Universit\`a di Lecce and Sezione INFN-Lecce, Lecce, Italy.
{\bf E-mail:} Luigi.Martina@le.infn.it
\\ $^\dag$ Feza G\"{u}rsey Institute, Istanbul,
Turkey and Department of Higher Mathematics, North Western State
Technical University, St. Petersburg, Russia. \\ {\bf E-mail:}
sheftel@gursey.gov.tr
\\ $^\ddag$ Centre de Recherches Math\'{e}matiques and D\'{e}partement de
math\'{e}matiques et statistique, Universit\'{e} de Montr\'{e}al,
Montr\'{e}al, Canada. \\ {\bf E-mail:} wintern@CRM.UMontreal.ca)
\vspace{3mm}
\\ Comments: 32 pages, Latex,
\\ submitted to {\it J. Phys. A: Math. Gen.} on April 13, 2001.
\\[1pt] Subj-class: Mathematical Physics; Exactly Solvable and
Integrable Systems
\\[1pt] MSC-class: 35Q75 (Primary) 57S20 (Secondary)
\begin{center}
\begin{abstract}
The main physical result of this paper are exact analytical
solutions of the heavenly equation, of importance in the general
theory of relativity. These solutions are not invariant under any
subgroup of the symmetry group of the equation. The main
mathematical result is a new method of obtaining noninvariant
solutions of partial differential equations with infinite
dimensional symmetry groups. The method involves the compatibility
of the given equations with a differential constraint, which is
automorphic under a specific symmetry subgroup, the latter acting
transitively on the submanifold of the common solutions. By
studying the integrability of the resulting conditions, one can
provide an explicit foliation of the entire solution manifold of
the considered equations.
\end{abstract}
\end{center}

\section{Introduction}
\label{sec:intro}

   An important problem for partial differential equations invariant with
respect to an infinite Lie group is to obtain non-invariant solutions that
admit no continuous symmetries of the equations. In our opinion, the old
approach of S. Lie \cite{lie} developed by Vessiot \cite{vessiot} and in
modern form by Ovsiannikov \cite{ovs}, which we call {\it group foliation},
is an adequate tool for treating this problem in the framework of Lie theory.
According to this method we foliate the solution space of the equations in
question into orbits, choosing for the foliation an infinite-dimensional
symmetry group.  Each orbit is determined by the {\it
automorphic system} joined to the original equations and considered as
invariant differential constraints. Due to the automorphic property of this
system, any of its solutions can be obtained from any other solution by a
transformation of the chosen symmetry subgroup.  This symmetry property makes
the automorphic system completely integrable if only one of its solutions can
be obtained.  The collection of orbits of all solutions of the original
equations is determined by the {\it resolving system}. Thus the problem
reduces to obtaining as many particular solutions of the resolving system as
possible.  Each of them will fix a particular automorphic system and the
corresponding orbit in the solution space of original equations.

  Group theory is usually used to obtain invariant
solutions. Here we show that it also provides a mechanism for obtaining
non-invariant solutions. We give examples of such solutions as
an application of the method.

  In this paper we further develop the method of group foliation by
introducing a procedure of {\it invariant integration}. It is used
for reconstructing the solution of the original equation corresponding to the
known particular solution of the resolving system. We apply the method for
obtaining non-invariant solutions of the `heavenly" equation
\begin{equation}
u_{xx}+u_{yy}=\kappa (e^u)_{tt}
\label{heav_xy}
\end{equation}
where $\kappa =\pm 1$ and the unknown $u$ depends on the time $t$ and two
space variables $x$ and $y$. Here and further subscripts of $u$ denote
partial differentiation with respect to corresponding variables.
This equation formally is a continuous
version of the Toda lattice \cite{mart}. It appears in various physical
theories, like the theory of area preserving diffeomorphisms \cite{bakas},
in the theory of the so-called gravitational instantons \cite{eguchi} and in
the general theory of relativity \cite{ward}. In this context it describes
self-dual Einstein spaces with Euclidean signature with one rotational
Killing vector.  Moreover it is a completely integrable system in the sense
of the existence of a Lax pair \cite{sav89,sav92}.

  The outline of the method is the following. We determine the total group
of point symmetries of the heavenly equation. For the group foliation we
choose its infinite subgroup of conformal transformations. We compute
differential invariants of this subgroup up to the second order inclusively
and obtain $5$ functionally independent differential invariants.
On account of the heavenly equation we are left with
$4$ invariants.  We choose three of them as new independent variables, the
same number as in the heavenly equation, and one is left for the new unknown.

 We obtain three first order {\it operators of invariant differentiation}
defined by the property that acting on a differential invariant they produce
again a differential invariant. These operators are determined by the
condition that they should commute with an arbitrary prolongation of any
element of the infinite symmetry Lie algebra chosen for the foliation.

 Extensive use of operators of invariant differentiation and their
commutator algebra for formulating the resolving system is a new
feature of the method suggested by one of the authors (M.B.S.) in
a recent article on the complex Monge-Amp\`ere equation \cite{ns}.
We derive the resolving system as a set of compatibility
conditions for the heavenly equation and its automorphic system,
using invariant cross-differentiation. Then we formulate the
resolving system in terms of the {\it commutator algebra of
operators of invariant differentiation} by discovering the fact
that this algebra together with its Jacobi identities, projected
on the solution manifold of the considered equation in the space
of differential invariants, is equivalent to the resolving system.

 We show how an Ansatz simplifying the commutator algebra of operators of
invariant differentiation leads to a particular class of solutions of the
resolving system. Then we use invariant integration to obtain the
corresponding solution of the heavenly equation and prove that this solution
is non-invariant.

\section{Lie group of point symmetries and\protect\\ differential invariants}
\label{sec:diffinvar}

\setcounter{equation}{0}

It is convenient to work with the heavenly equation using the complex
coordinates $z=(x+iy)/2, \;\bar z=(x-iy)/2$
\begin{equation}
u_{z\bar z}=\kappa (e^u)_{tt}.
\label{heav}
\end{equation}
A standard calculation of the total symmetry group of the heavenly equation
gives the following result for the symmetry generators of all one-parameter
subgroups \cite{soliani}
\begin{eqnarray}
T \!\!&=&\!\! \partial_t, \qquad G=t\partial_t+2\partial_u, \nonumber
\\ X_a \!\!&=&\!\! a(z)\partial_z +\bar a(\bar z)\partial_{\bar z}
-(a^\prime(z)+\bar a^\prime(\bar z))\partial_u ,
\label{symgen}
\end{eqnarray}
where $T$ is the generator of translations in $t$, $G$ is the generator of
a dilation of time accompanied by a shift of $u$: $t=\tilde t e^\tau,\;
u=\tilde u +2\tau$ and $X_a$ is a generator of the conformal transformations
\begin{equation}
z=\phi(\tilde z),\quad \bar z=\bar \phi(\tilde{\bar
z}),\quad u(z,\bar z,t) = \tilde u(\tilde z,\tilde{\bar z},t)
-\ln\bigl(\phi^\prime(\tilde z){\bar\phi}^\prime(\tilde{\bar z})\bigr),
\label{conform}
\end{equation}
where $a(z)$ and $\phi(z)$ are arbitrary holomorphic functions of $z$ (see
also \cite{boywin}).

 The Lie algebra of the symmetry generators is determined by the commutation
relations
\begin{equation}
[T,G]=T,\quad [T,X_a]=0,\quad [G,X_a]=0,\quad
[X_a,X_b]=X_{ab^\prime-ba^\prime} ,
\label{algeb}
\end{equation}
which show that the generators $X_a$ of conformal transformations form
an infinite-dimensional subalgebra.

  We choose for the group foliation the corresponding infinite symmetry
subgroup of all holomorphic transformations in $z$, {\it i.e.} the conformal
group. Differential invariants of this group are the invariants of all its
generators $X_a$ of the form (\ref{symgen}) in the prolongation spaces. This
means that they can depend on independent variables, the unknowns and also on
the partial derivatives of the unknowns allowed by the order of the
prolongation.  The {\it order} $N$ of the differential invariant is defined
as the order of the highest derivative which this invariant
depends on.  The determining equation for differential invariants $\Phi$ of
the order $N\le 2$ has the form
\begin{equation}
\X2(\Phi)=0 ,
\label{detinv}
\end{equation}
where $\X2$ is the second prolongation of the generator $X_a$ (\ref{symgen})
of the conformal group defined by the standard prolongation formulae
\begin{eqnarray}
\X2\!\!&\!\!=\!\!&\!\!a\partial_z+\bar a\partial_{\bar z}
-\left(a^\prime+{\bar a}^\prime\right)\partial_u
-\left(a^{\prime\prime}+a^\prime u_z\right)\partial_{u_z}
-\left({\bar a}^{\prime\prime}+{\bar a}^\prime u_{\bar z}\right)
\partial_{u_{\bar z}}\nonumber
\\ \!\!&\!\! \!\!&\!\! \mbox{}-\left(a^{\prime\prime\prime}
+a^{\prime\prime}u_z+2a^\prime u_{zz}\right)\partial_{u_{zz}}
-\left({\bar a}^{\prime\prime\prime}
+{\bar a}^{\prime\prime}u_{\bar z}+2{\bar a}^\prime u_{\bar z\bar z}\right)
\partial_{u_{\bar z\bar z}}\nonumber
\\ \!\!&\!\! \!\!&\!\! \mbox{}-a^\prime u_{zt}\partial_{u_{zt}}
-{\bar a}^\prime u_{\bar zt}\partial_{u_{\bar zt}}
-\left(a^\prime+{\bar a}^\prime\right)u_{z\bar z}\partial_{u_{z\bar z}},
\label{X2}
\end{eqnarray}
where $a=a(z)$ and $\bar a=\bar a(\bar z)$.

 The integration of eq.(\ref{detinv}) gives $5$ functionally independent
differential invariants up to the second order inclusively
\begin{equation}
t,\qquad u_t,\qquad u_{tt},\qquad \rho=e^{-u}u_{z\bar z},\qquad
\eta=e^{-u}u_{zt}u_{\bar zt}
\label{difinv}
\end{equation}
and all of them turn out to be real.
This allows us to express the heavenly equation (\ref{heav}) solely in terms
of the differential invariants
\begin{equation}
u_{tt}=\kappa\rho-u_t^2 .
\label{heavinv}
\end{equation}

\section{Operators of invariant differentiation and a basis of differential
invariants}
\label{sec:invdiff}

\setcounter{equation}{0}

  {\it Operators of invariant differentiation} are linear
combinations of total derivative operators with respect to independent
variables. Their coefficients depend on local coordinates of the
prolongation space. They are defined by the special property that, acting on
any (differential) invariant, they map it again into a differential
invariant.  Being first-order differential operators, they raise the order of
a differential invariant by one. Invariance requires that these differential
operators commute with any infinitely prolonged generator $X_a$
(\ref{symgen}) of the conformal symmetry group. It is obvious (see
\cite{ovs} par. 24.2 for a complete proof) that the total
number of independent operators of invariant differentiation is equal to the
number of total derivative operators, that is to the number of independent
variables (which is three in the present case).

  We look for operators of invariant differentiation in the form
\begin{equation}
\delta=\lambda_1D_t+\lambda_2D_z+\lambda_3D_{\bar z}=\sum_{i=1}^{3}\lambda_i
D_i
\label{invdifform}
\end{equation}
where $D_1=D_t,\; D_2=D_z,\; D_3=D_{\bar z}$ are operators of total
derivatives with respect to the subscripts. We look for the coefficients
$\lambda_i$ satisfying  the condition of commutativity of $\delta$ with the
infinite prolongation $\X\infty$ of the generator $X_a$ (\ref{symgen}). It
can be decomposed as the sum of the infinite prolongation of the symmetry
generator in the evolution form $\widehat X$ \cite{olv} and the linear
combination of the total derivative operators
\begin{equation}
\X\infty= \widehat X +\sum_{j=1}^{3}\xi^jD_j=\widehat X +a(z)D_z +\bar a(z)
D_{\bar z},
\label{decomp}
\end{equation}
where from the form of $X_a$ we take
\begin{equation}
\xi^1=\xi^t=0,\quad \xi^2=\xi^z=a(z),\quad
\xi^3=\xi^{\bar z}=\bar a(\bar z).
\label{xi^i}
\end{equation}
The generator $\widehat X$ in the
evolution form commutes with all the total derivatives $D_j$:
$\Bigl[D_i,\widehat X\Bigr]=0$ and hence we have the standard commutation
relation
\begin{equation}
\Bigl[D_i,\X\infty\Bigr]=\Bigl[D_i,\sum_{j=1}^{3}\xi^jD_j\Bigr]
=\sum_{j=1}^{3}D_i(\xi^j)D_j.
\label{standcom}
\end{equation}
We use it in the determining relation for operators of invariant
differentiation
\begin{eqnarray}
&
&\Bigl[\delta,\X\infty\Bigr]=\sum_{i}^{}\Bigl[\lambda_iD_i,\X\infty\Bigr]
=\sum_{i}^{}\Bigl(\sum_{j}\lambda_iD_i[\xi^j]D_j-\X\infty(\lambda_i)D_i\Bigr)
\nonumber
\\ & &=\sum_{i}^{}\Bigl(\sum_{j}\lambda_jD_j[\xi^i]-\X\infty(\lambda_i)\Bigr)D_i=0.
\label{comdelt}
\end{eqnarray}
The final equation for the coefficients $\lambda_i$ of the
operators of invariant differentiation (see eq.(24.2.3) of \cite{ovs}) is
\begin{equation}
\X\infty(\lambda_i)=\sum_{j=1}^{3}\lambda_jD_j[\xi^i].
\label{eqinvdif}
\end{equation}
Using (\ref{xi^i}) and restricting ourselves to the
second prolongation $\X2$ of the symmetry generator, the equation
(\ref{eqinvdif}) leads to
\begin{equation}
\X2(\lambda_1)=0,\quad \X2(\lambda_2)=\lambda_2 a^\prime(z), \quad
\X2(\lambda_3)=\lambda_3 {\bar a}^\prime(\bar z)
\label{eqlamb}
\end{equation}
where primes denote derivatives and
\[\lambda_i=\lambda_i(t,z,\bar z,u,u_t,u_z,
u_{\bar z},u_{zt},u_{\bar zt},u_{z\bar z},u_{zz},u_{\bar z\bar z}).\]
Here $\X2$ is the second prolongation of the generator $X_a$ of the conformal
group defined by eq.(\ref{X2}).

  Equations (\ref{eqlamb}) are easily solved by the method of characteristics
and we choose $3$ simplest linearly independent solutions for the
coefficients $\lambda_i$ of the three operators of invariant differentiation
\begin{eqnarray}
& \lambda^1_1=1,\; \lambda^1_2=0,\; \lambda^1_3=0,\quad
\lambda^2_1=0,\; \lambda^2_2=e^{-u}u_{\bar zt},\; \lambda^2_3=0,&\nonumber
\\ &\lambda^3_1=0,\; \lambda^3_2=0,\; \lambda^3_3=e^{-u}u_{zt}.&
\label{sollamb}
\end{eqnarray}
From here we obtain a basis for the operators of invariant differentiation
\begin{equation}
\delta=D_t,\qquad \Delta=e^{-u}u_{\bar zt}D_z,\qquad
\bar \Delta=e^{-u}u_{zt}D_{\bar z}.
\label{deltas}
\end{equation}

 The {\it basis of differential invariants} is defined as a minimal finite
set of invariants of a symmetry group from which any other differential
invariant of this group can be obtained by a finite number of invariant
differentiations and operations of taking composite functions. The proof
of the existence and finiteness of the basis was given by Tresse \cite{tresse}
and in a more modern form by Ovsiannikov \cite{ovs}.

  In our example the basis of differential invariants is formed by the set of
three invariants $t,u_t,\rho$, while two other invariants $u_{tt}$ and $\eta$
of eq.(\ref{difinv}) are given by the relations
\begin{equation}
u_{tt}=\delta(u_t),\qquad \eta\equiv e^{-u}u_{zt}u_{\bar zt}=\Delta(u_t)
=\bar\Delta(u_t).
\label{invdepend}
\end{equation}
All other functionally independent higher-order invariants can be obtained by
acting with operators of invariant differentiation on the basis
$\{t,u_t,\rho\}$.  In particular, the following third-order invariants
generated from the 2nd-order invariant $\rho$ by invariant differentiations
will be involved in our construction
\begin{equation}
\sigma=\Delta(\rho),\quad \bar\sigma=\bar\Delta(\rho),\quad \tau=\delta(\rho)
\equiv \rho_t.
\label{sigtau}
\end{equation}

 The operators of invariant differentiation form the commutator algebra
\begin{eqnarray}
& {\displaystyle
\bigl[\delta,\Delta\bigr]=\left(\kappa\,\frac{\bar\sigma}{\eta}-3u_t\right)
\Delta,\qquad
\bigl[\delta,\bar\Delta\bigr]=\left(\kappa\,\frac{\sigma}{\eta}-3u_t\right)
\bar\Delta }&\nonumber
\\  & {\displaystyle
\bigl[\Delta,\bar\Delta\bigr]=\left(\frac{\Delta(\eta)}{\eta}-(u_t\rho+\tau)
\right)\bar\Delta -\left(\frac{\bar\Delta(\eta)}{\eta}-(u_t\rho+\tau)\right)
\Delta } &
\label{comalg}
\end{eqnarray}
which form a Lie algebra over the field of invariants of the conformal group,
in agreement with Ovsiannikov's lemma 24.2 \cite{ovs}.

  The commutator algebra is simplified by introducing two new operators of
invariant differentiation $Y$ and $\bar Y$ instead of $\Delta$ and
$\bar\Delta$ and two new variables $\lambda$ and $\bar\lambda$ instead of
$\sigma$ and $\bar\sigma$, defined by
\begin{equation}
\Delta=\eta Y,\quad \bar\Delta=\eta\bar Y,\quad \sigma=\eta\lambda,\quad
\bar\sigma=\eta\bar\lambda ,
\label{Ydef}
\end{equation}
and becomes
\begin{eqnarray}
& {\displaystyle \bigl[\delta,Y\bigr]=\left(\kappa\bar\lambda
-3u_t-\frac{\delta(\eta)}{\eta}\right) Y, \qquad \bigl[\delta,\bar
Y\bigr]=\left(\kappa\lambda-3u_t-\frac{\delta(\eta)}{\eta}\right)\bar Y,\quad
}& \nonumber
\\&\bigl[Y,\bar Y\bigr]={\displaystyle\frac{(u_t\rho+\tau)}{\eta}}
\left(Y - \bar Y\right).&
\label{Yalg}
\end{eqnarray}
Equations (\ref{invdepend}) and (\ref{sigtau}) imply the following properties
of the operators $Y$ and $\bar Y$
\begin{equation}
Y(u_t)=\bar Y(u_t)=1,\quad Y(\rho)=\lambda,\quad \bar Y(\rho)=\bar\lambda.
\label{Yact}
\end{equation}

\section{Automorphic and resolving equations}
\label{sec:autresys}

\setcounter{equation}{0}

 We have four independent differential invariants $t,u_t,\rho,\eta$
on the solution manifold of the heavenly equation (\ref{heavinv}).
We choose three of them $t,u_t,\rho$ as
new invariant independent variables, the same number as in the original
equation (\ref{heav}), and consider the fourth one $\eta$ as a real function
$F$ of these three
\begin{equation}
\eta=F(t,u_t,\rho)\quad\iff\quad
u_{zt}u_{\bar zt}e^{-u}=F\left(t,u_t,u_{z\bar z}e^{-u}\right),
\label{autom}
\end{equation}
which gives us the general form of the automorphic equation, {\it i.e.}
invariant differential constraint.

  Our next task is to derive the {\it resolving equations} for the heavenly
equation. This will account for all integrability conditions of the system
(\ref{heavinv}), (\ref{autom}) in an explicitly invariant form. If we pick
a particular solution of this {\it resolving system} for $F$ and use it in
the right-hand side of (\ref{autom}), then the latter equation will possess
the automorphic property: each solution of it can be obtained from any other
solution by an appropriate conformal symmetry transformation.

  We consider the automorphic equation (\ref{autom}) divided by $F$ in the
form
\begin{equation}
Y(u_t)=1 ,
\label{Yautom}
\end{equation}
and the heavenly equation (\ref{heavinv}) in the form
\begin{equation}
\delta(u_t)=\kappa\rho-u_t^2.
\label{deltheav}
\end{equation}
We put $\eta = F$ in the definitions (\ref{Ydef}) of $Y$ and $\bar Y$ and
in their commutation relations (\ref{Yalg}).
The integrability condition for the system (\ref{Yautom}) and (\ref{deltheav})
is obtained by the {\it invariant cross-differentiation} with $\delta$ and
$Y$ with the use of their commutation relation (\ref{Yalg})
\begin{equation}
\delta(F)=\bigl[\kappa(\lambda+\bar\lambda)-5u_t\bigr]F.
\label{1}
\end{equation}
Since this equation involves $\lambda$ and $\bar\lambda$ we use their
definitions in eq.(\ref{Yact})
\begin{equation}
Y(\rho)=\lambda,\qquad \bar Y(\rho)=\bar\lambda
\label{lambdef}
\end{equation}
and obtain the integrability condition for these two equations by
the invariant cross-differentiation by $\bar Y$ and $Y$ using their
commutation relation from eq.(\ref{Yalg})
\begin{equation}
F\bigl(Y(\bar\lambda)-\bar
Y(\lambda)\bigr)=(u_t\rho+\tau)(\lambda-\bar\lambda).
\label{3}
\end{equation}
This equation contains $\tau$, so we use its definition (\ref{sigtau})
\begin{equation}
\delta(\rho)=\tau .
\label{taudef}
\end{equation}
Using the invariant cross-differentiation with $Y$ or $\bar Y$ and
$\delta$, we obtain the compatibility conditions of eq.(\ref{taudef}) with
each of equations (\ref{lambdef})
\begin{equation}
\delta(\lambda)=Y(\tau)+2u_t\lambda-\kappa\lambda^2
\label{2}
\end{equation}
and
\begin{equation}
\delta(\bar\lambda)=\bar Y(\tau)+2u_t\bar\lambda-\kappa{\bar\lambda}^2 .
\label{b2}
\end{equation}
These are complex conjugate to each other. There is one more differential
consequence of the obtained resolving equations. This is the integrability
condition of the equation (\ref{3}) solved with respect to $Y(\bar\lambda)$
together with the equation (\ref{b2}). It is obtained by the invariant
cross-differentiation of these equations by $\delta$ and $Y$. Using the
other resolving equations it can be brought to the form
\begin{eqnarray}
 & & F\bigl(Y(\bar\lambda)+\bar Y(\lambda)\bigr)=
-(u_t\rho+\tau)(\lambda+\bar\lambda)\nonumber
\\ & & \mbox{}+2\kappa\bigl[\delta(\tau)+2F+4u_t\tau+\kappa\rho^2+2u_t^2\rho
\bigr].
\label{4}
\end{eqnarray}

 The resolving equations (\ref{1}), (\ref{3}), (\ref{2}), (\ref{b2}) and
(\ref{4}) form a closed resolving system if we assume that not only the
2nd-order differential invariant $\eta=F$, but also the 3rd-order
differential invariants $\lambda,\bar\lambda$ and $\tau$ are functions of
$t,u_t,\rho$. They should be regarded as additional unknowns in these
equations, so the resolving system consists of $5$ partial differential
equations with $4$ unknowns $F,\lambda,\bar\lambda$ and $\tau$. The operators
of invariant differentiation are projected on the solution manifold of the
heavenly equation and on the space of differential invariants treated as new
independent variables. We keep the same notation for the projected operators
of invariant differentiation and write them in the form
\begin{equation}
\delta=\partial_t+(\kappa\rho-u_t^2)\partial_{u_t}+\tau\partial_\rho,
\quad Y=\partial_{u_t}+\lambda\partial_\rho,\quad
\bar Y=\partial_{u_t}+\bar\lambda\partial_\rho .
\label{deltproj}
\end{equation}
Here we have used the following properties of these operators
\begin{eqnarray}
 & \delta(t)=1,\quad \delta(u_t)=\kappa\rho-u_t^2,\quad \delta(\rho)=\tau &
\label{deltprop}
\\ & Y(t)=\bar Y(t)=0,\quad
Y(u_t)=\bar Y(u_t)=1,\quad Y(\rho)=\lambda,\quad \bar Y(\rho)=\bar\lambda, &
\nonumber
\end{eqnarray}
which follow from their definitions, equations (\ref{invdepend}),
(\ref{sigtau}), (\ref{Yact}) and the heavenly equation in the form
(\ref{deltheav}). If we used for the operators of invariant differentiation
$\delta,Y,\bar Y$ the formulae (\ref{deltproj}) in the resolving equations
(\ref{1}), (\ref{3}), (\ref{2}), (\ref{b2}) and (\ref{4}), then we would
obtain the resolving system in an explicit form as a system of $5$
first-order PDEs with $4$ unknowns $F,\lambda,\bar\lambda,\tau$ and $3$
independent variables $t,u_t,\rho$. This system is
passive, {\it i.e.} it has no further algebraically independent first-order
integrability conditions.

  The commutator relations (\ref{Yalg}) were satisfied identically by
the operators of invariant differentiation. On the contrary, for the
projected operators (\ref{deltproj}) these commutation relations and even
the Jacobi identity
\begin{equation}
\bigl[\delta,[Y,\bar Y]\bigr]+\bigl[Y,[\bar Y,\delta]\bigr]
+\bigl[\bar Y,[\delta,Y]\bigr]=0
\label{jacobi}
\end{equation}
are not identically satisfied, but only on account of the resolving
equations. It is easy to check that even a stronger statement is valid.
\begin{theorem}
The commutator algebra\/ {\rm (\ref{Yalg})} of the operators
of invariant differentiation $\delta,Y,\bar Y$, together with the Jacobi
identity\/ {\rm(\ref{jacobi})}, is equivalent to the
resolving system for the heavenly equation and hence provides a commutator
representation for this system.
\end{theorem}
This theorem means that the complete set of the resolving equations is
encoded in the commutator algebra of the operators of invariant
differentiation and provides the easiest way to derive the resolving system.
In Section $6$ we shall show how the commutator representation of the
resolving system can lead to a useful Ansatz for solving this system.

\section{Invariant and non-invariant solutions}
\label{sec:invsol}

\setcounter{equation}{0}

  {\it Invariant solutions} are defined as solutions that are invariant
with respect to a symmetry subgroup of the equation. {\it Non-invariant
solutions} are those solutions which are not invariant with respect to any
one-parameter symmetry group of the equation. We present here a simple
derivation of the infinitesimal criterion of invariance of solutions.

  Consider a general form of the generator of a one-parameter symmetry of
the heavenly equation as a linear combination of symmetry generators
(\ref{symgen}) with arbitrary real constant coefficients $\alpha$ and $\beta$
\begin{equation}
X= \alpha\partial_t + \beta\left(t\partial_t+2\partial_u\right)
 + a(z)\partial_z + \bar a(\bar z)\partial_{\bar z}
-\left(a^\prime(z)+\bar a^\prime(\bar z)\right)\partial_u
\label{symmet}
\end{equation}
where $a(z)$ is an arbitrary holomorphic function. The infinitesimal
criterion for the invariance of the solution $u=f(z,\bar z,t)$ with respect
to the generator $X$ has the general form (see par. 19.2.1 of \cite{ovs})
\begin{equation}
X(f-u)|_{u=f}=0 ,
\label{gencrit}
\end{equation}
which for $X$ defined by the formula (\ref{symmet}) becomes
\begin{equation}
(\alpha+\beta t)f_t + a(z)f_z + \bar a(\bar z)f_{\bar z} =
 2\beta - a^\prime(z) - \bar a^\prime(\bar z).
\label{critinv}
\end{equation}
The invariance criterion can be summed up as follows.
\begin{proposition}
If there exists a holomorphic function $a(z)$ and constants $\alpha$ and
$\beta$, not all equal to zero, such that the equation (\ref{critinv}) is
satisfied, then the solution $u=f(z,\bar z,t)$ is invariant. Otherwise this
solution is non-invariant.
\end{proposition}

  From this proposition one can derive some criteria for the
non-invariance of solutions. For example, we consider the case when
$\alpha=0$ and $\beta=0$ so that equation (\ref{critinv}) is a criterion of
conformal invariance. The general solution of eq.(\ref{critinv}) in this
case has the form
\begin{equation}
u = \ln{f(\xi,t)} -\ln{a(z)} -\ln{\bar a(\bar z)}
\label{invsol}
\end{equation}
where
\begin{equation}
\xi = i \left(\int\frac{dz}{a(z)} - \int\frac{d\bar z}{\bar a(\bar z)}
\right).
\label{xi}
\end{equation}
The invariant $\rho$ defined by eq.(\ref{difinv}) becomes
\begin{equation}
\rho = \frac{ff_{\xi\xi}-f_\xi^2}{f^3}
\label{ro}
\end{equation}
and the invariants $\sigma$ and $\bar\sigma$, defined by eqs.(\ref{sigtau})
as
\[\sigma = e^{-u} u_{\bar zt} D_z(\rho),\qquad
\bar\sigma = e^{-u} u_{zt} D_{\bar z}(\rho) ,\]
are equal to each other
\begin{equation}
\bar\sigma = \sigma = \frac{(ff_{t\xi}-f_tf_\xi)}{f^3}\times
\left(\frac{ff_{\xi\xi}-f_\xi^2}{f^3}\right)_\xi ,
\label{sigm}
\end{equation}
where the subscripts denote partial differentiations. Hence the
necessary condition for a solution to be conformally invariant is the
equality
\begin{equation}
\bar\sigma = \sigma \quad (\iff \bar\lambda = \lambda).
\label{confinv}
\end{equation}
The converse statement gives the criterion for a solution to be conformally
non-invariant.
\begin{corollary}
The sufficient condition for a solution of the heavenly equation to be
conformally non-invariant is that the following inequality should be
satisfied
\begin{equation}
\bar\sigma \ne \sigma
\label{noninv}
\end{equation}
\end{corollary}
(or equivalently $\bar\lambda \ne \lambda$).

 Concerning the practical use of this statement we must remark that even if
the inequality (\ref{noninv}) is satisfied for a solution of the resolving
system it could become the equality (\ref{confinv}) on the corresponding
solution of the heavenly equation.  Nevertheless the above criterion is
useful, meaning that we should avoid solutions of the resolving equations
satisfying eq.(\ref{confinv}) in order not to end up with conformally
invariant solutions.

\section{Particular solutions of the resolving\protect\newline system}
\label{sec:resolv}

\setcounter{equation}{0}

  Here we show that the commutator representation of the resolving system
can prompt Ansatzes, leading to particular solutions of the resolving
equations. Attempts to
solve the commutation relations by imposing relations between the
operators of invariant differentiation lead to invariant solutions of the
heavenly equation. This is the case with the Ansatz $\bar Y = Y$. Then
the expressions (\ref{deltproj}) for $Y,\bar Y$ imply $\bar\lambda =
\lambda$, so the condition (\ref{noninv}) of Corollary~\thecorollary\ is not
satisfied. Hence we obtain a conformally invariant solution of the heavenly
equation.

 Another possible simplifying Ansatz is that the operators $Y$ and $\bar Y$
commute and we have
\begin{equation}
\tau = -u_t \rho \quad \Rightarrow \quad [Y,\bar Y] = 0 ,
\label{ans}
\end{equation}
but $\bar Y\ne Y$.

  Before solving the resolving system with the Ansatz (\ref{ans}) we keep in
mind that $F\ne 0$. Indeed the case $F=0$ is singular for the
derivation of the resolving equations and should be treated separately.
We shall consider first the case $F=0$ and show that it leads to invariant
solutions of the heavenly equation.

  Putting $F=0$ in equation (\ref{autom}) we obtain
\begin{equation}
u_{zt} = 0, \qquad u_{\bar zt} =0
\label{f0}
\end{equation}
and hence we have the separation
\begin{equation}
u = \alpha(t) + \beta(z,\bar z).
\label{sepu}
\end{equation}
Substituting this expression for $u$ into the heavenly equation (\ref{heav})
we obtain
\begin{equation}
e^{\alpha(t)}\left(\alpha^{\prime\prime}(t)+\left(\alpha^\prime(t)\right)^2
\right)=\kappa e^{-\beta(z,\bar z)}\beta_{z\bar z}(z,\bar z)=2l ,
\label{sepheav}
\end{equation}
where $l=\bar l$ is a separation constant and the primes denote derivatives
in $t$. Integrating the equation for $\alpha$ we obtain
\begin{equation}
\alpha(t)=\ln{\left(lt^2+C_1t+C_2\right)}
\label{alp}
\end{equation}
where $C_1,C_2$ are arbitrary real constants.
The equation for $\beta(z,\bar z)$
\begin{equation}
\beta_{z\bar z}=2\kappa l e^{\beta}
\label{liouv}
\end{equation}
is the Liouville equation if $\kappa=1$ and the
\`{}$\!$pseudo-Liouville$\!$\'{}
equation for $\kappa=-1$. Its general solution is
\begin{equation}
\beta(z,\bar z) = \ln{a^\prime(z)} + \ln{{\bar a}^\prime(\bar z)}
-2 \ln{\left(a(z)+\bar a(\bar z)\right)} -\ln{l}
\label{plus}
\end{equation}
if $\kappa=1$ and
\begin{equation}
\beta(z,\bar z) = \ln{a^\prime(z)} + \ln{{\bar a}^\prime(\bar z)}
-2 \ln{\left(a(z)\bar a(\bar z)+1\right)} -\ln{l}
\label{minus}
\end{equation}
if $\kappa=-1$. Here $a(z)$ is an arbitrary holomorphic function and the
primes denote derivatives.
Thus, the corresponding solutions of the heavenly equation
are given by the equation (\ref{sepu}) with $\alpha(t)$ determined by the
formula (\ref{alp}) and $\beta(z,\bar z)$ determined by the formula
(\ref{plus}) or (\ref{minus}).

  To obtain the simplest representative of the orbit of
solutions we apply simplifying symmetry transformations:
the conformal transformation
\[a(z)\mapsto z,\qquad \bar a(\bar z)\mapsto \bar z , \]
the suitable time translation and the dilation of time accompanied by a
shift of $u$
\[u\mapsto u+\ln{l},\qquad t\mapsto \frac{t}{\sqrt{l}}. \]
The resulting solutions become
\begin{equation}
u = \ln{\left(t^2+C\right)} -2\ln{(z+\bar z)}\qquad {\rm if}\quad \kappa=1 ,
\label{f0simplus}
\end{equation}
\begin{equation}
u = \ln{\left(t^2+C\right)} -2\ln{(z\bar z+1)}\qquad {\rm if}\quad \kappa=-1,
\label{f0siminus}
\end{equation}
where $C$ is an arbitrary real constant.

  To perform a check of invariance of the solutions (\ref{f0simplus}) and
(\ref{f0siminus}), we substitute them into the criterion of invariance
(\ref{critinv}) and make a splitting in $t$. Then we obtain $\alpha=0$ and
if $C\ne 0$, then also $\beta=0$. For $C=0$ the constant $\beta$ can be
arbitrary. We also obtain a differential equation for $a(z)$ and $\bar
a(\bar z)$
\begin{equation}
a^\prime(z)+{\bar a}^\prime(\bar z)=2\,\frac{a(z)+\bar a(\bar z)}{z+\bar z}
\quad {\rm for}\quad \kappa=1
\label{odeplus}
\end{equation}
and
\begin{equation}
a^\prime(z)+{\bar a}^\prime(\bar z)=2\,\frac{\bar z a(z)+ z\bar a(\bar
z)}{z\bar z+1}
\quad {\rm for}\quad \kappa=-1 ,
\label{odemin}
\end{equation}
with the trivial solutions
\begin{equation}
a=i,\; \bar a=-i \quad {\rm for}\; \kappa=1;\qquad
a=iz,\; \bar a=-iz \quad {\rm for}\; \kappa=-1.
\label{a(z)}
\end{equation}
Thus, we have proved that there exist $\alpha,\beta$ and $a(z),\bar a(\bar z)$
such that the criterion (\ref{critinv}) of invariance of solutions is
satisfied for our solutions (\ref{f0simplus}) and (\ref{f0siminus}). Hence
the case $F=0$ corresponds to invariant solutions.

  In the following we assume that $F\ne 0$ and consider the resolving
equations with the Ansatz (\ref{ans}). Equations (\ref{3}) and (\ref{4})
become respectively
\[Y(\bar\lambda)=\bar Y(\lambda)\quad {\rm and}\quad
Y(\bar\lambda) + \bar Y(\lambda) = 4\kappa \]
and hence
\begin{equation}
\bar Y(\lambda) = 2\kappa,\qquad  Y(\bar\lambda) = 2\kappa.
\label{Ylamb}
\end{equation}
Next we consider the compatibility condition of the system of equations
(\ref{2}), (\ref{b2}) and the first equation in (\ref{Ylamb}). Because of the
formula (\ref{ans}) the first equation becomes
\begin{equation}
\delta(\lambda) = u_t\lambda -\kappa\lambda^2 -\rho .
\label{2mod}
\end{equation}
Then, using cross-differentiation of the
invariant operators  $\delta$ and $\bar Y$,
their commutator (\ref{Yalg}) and eq.(\ref{Ylamb}), we obtain a very simple
result
\begin{equation}
\lambda + \bar\lambda = 2\kappa u_t.
\label{labla}
\end{equation}
Solving this equation with respect to $\bar\lambda$, substituting
in equation (\ref{b2}) and using the equation (\ref{2mod}) to express
$\delta(\lambda)$, we obtain a quadratic equation for $\lambda$
\[ \lambda^2 -2\kappa u_t\lambda +2\kappa\rho = 0 ,\]
with the solution
\begin{equation}
\lambda = \kappa u_t + i\sqrt{2\kappa\rho -u_t^2} ,
\label{lamb}
\end{equation}
where we have chosen the $+$ sign before the square root. Equation
(\ref{labla}) gives the result
\begin{equation}
\bar\lambda = \kappa u_t - i\sqrt{2\kappa\rho -u_t^2}
\label{blamb}
\end{equation}
which is complex conjugate to (\ref{lamb}) provided the condition
\begin{equation}
2\kappa\rho -u_t^2 \ge 0
\label{non-neg}
\end{equation}
is satisfied. The obtained expressions for $\lambda$ and
$\bar\lambda$ satisfy the equations (\ref{Ylamb}), (\ref{2mod}) and
its complex conjugate, and hence all the resolving equations apart from
the equation (\ref{1}). We rewrite this last equation for the new unknown
$f=\ln{F}$ as
\begin{equation}
f_t + (\kappa\rho -u_t^2) f_{u_t} -u_t\rho f_\rho = -3u_t
\label{1mod}
\end{equation}
and solve it by the method of characteristics obtaining the general
solution
of the equation (\ref{1})
\begin{equation}
F = \rho^3 \varphi(\xi,\theta)\quad {\rm where}\quad
\xi=\frac{2\kappa\rho-u_t^2}{\rho^2}, \quad
\theta=t-\frac{\kappa}{\rho}\,\left(u_t+\sqrt{2\kappa\rho -u_t^2}\right),
\label{sol1}
\end{equation}
where $\varphi$ is an arbitrary real differentiable function.

 Finally we sum up our results for the particular solution of the resolving
system which follows from our Ansatz (\ref{ans})
\begin{equation}
F = \rho^3 \varphi(\xi,\theta),\quad \tau = -u_t\rho,\quad
\lambda = \kappa u_t + i\sqrt{2\kappa\rho -u_t^2}, \quad
\bar\lambda = \kappa u_t - i\sqrt{2\kappa\rho -u_t^2}
\label{solvresolv}
\end{equation}
This will be used in the next section for obtaining the corresponding
solution of the heavenly equation. Since $\bar\lambda\ne\lambda$ the
condition (\ref{noninv}) of Corollary \thecorollary\ for non-invariance
of this solution is satisfied.

\section{Invariant integration and non-invariant \protect\newline
         solution of the heavenly equation}
\label{sec:non-invsol}

\setcounter{equation}{0}

  In this section we reconstruct the solution of the heavenly equation
starting from the particular solution (\ref{solvresolv}) of the resolving
system. We demonstrate here the procedure of {\it invariant integration}
which amounts to the transformation of equations to the form of the exact
invariant derivative. Then we drop the operator of invariant differentiation
adding the term playing the role of the integration constant
which is an arbitrary element of the kernel of this operator.

  We start from our Ansatz (\ref{ans}) using the definitions
$\tau=\delta(\rho)$ and $\delta=D_t$
\begin{equation}
D_t(\ln{\rho}) = D_t(-u).
\label{ansmod}
\end{equation}
We integrate this equation in the form
\[ \ln{\rho} = -u +\ln{\gamma_{z\bar z}(z,\bar z)},\]
where the last term is a function to be determined. Solving this equation
with respect to $\rho$ and using the definition of $\rho$ we obtain
\[ \rho = e^{-u} u_{z\bar z} = e^{-u}\gamma_{z\bar z}(z,\bar z) \]
and hence $u_{z\bar z} = \gamma_{z\bar z}(z,\bar z)$. This implies the
following form of the solution
\begin{equation}
u(z,\bar z,t) =
\gamma(z,\bar z) + \alpha(z,t) +\bar\alpha(\bar z,t),
\label{solform}
\end{equation}
where $\gamma,\alpha$ and $\bar\alpha$ are arbitrary smooth functions of two
variables. After the substitution of this expression into the heavenly
equation (\ref{heav}) it becomes
\begin{equation}
e^{\alpha(z,t)+\bar\alpha(\bar z,t)}\left[\alpha_{tt}(z,t)
+\bar\alpha_{tt}(\bar z,t)+\bigl(\alpha_t(z,t)
+\bar\alpha_t(\bar z,t)\bigr)^2\right] = \kappa e^{-\gamma(z,\bar z)}
\gamma_{z\bar z}(z,\bar z).
\label{heavmod}
\end{equation}

  Next we rewrite the formulae (\ref{solvresolv}) for $\lambda$ and
$\bar\lambda$ in the form of exact invariant derivatives
\begin{equation}
Y(\sqrt{2\kappa\rho-u_t^2}-i\kappa u_t) = 0,\quad
\bar Y(\sqrt{2\kappa\rho-u_t^2}+i\kappa u_t) = 0.
\label{YbYeq}
\end{equation}
On account of the definitions (\ref{Ydef}),
the operators $Y$ and $\bar Y$ can be written as
\[ Y=\frac{1}{F}\,\Delta=\frac{e^{-u}u_{\bar zt}}{F}\,D_z,\quad
\bar Y=\frac{1}{F}\,\bar\Delta=\frac{e^{-u}u_{zt}}{F}\,D_{\bar z} \]
and the equations (\ref{YbYeq}) become
\[(\sqrt{2\kappa\rho-u_t^2}+i\kappa u_t)_{\bar z} = 0,\quad
(\sqrt{2\kappa\rho-u_t^2}-i\kappa u_t)_z = 0. \]
They are integrated in the form
\begin{equation}
\sqrt{2\kappa\rho-u_t^2}+i\kappa u_t=\psi(z,t),\quad
\sqrt{2\kappa\rho-u_t^2}-i\kappa u_t=\bar\psi(\bar z,t),
\label{intYeq}
\end{equation}
where $\psi,\bar\psi$ are arbitrary smooth functions.
Taking the difference of two equations (\ref{intYeq}) we obtain
\[u_t=-\frac{i\kappa}{2}\,[\psi(z,t)-\bar\psi(\bar z,t)]
= \alpha_t(z,t) +\bar\alpha_t(\bar z,t), \]
where the last equality follows from the expression (\ref{solform}) for $u$.
Separation of $z,\bar z$ in the last equality leads to
\[ \alpha_t(z,t)+\frac{i\kappa}{2}\,\psi(z,t) =
-\left[\bar\alpha_t(\bar z,t)-\frac{i\kappa}{2}\,\bar\psi(\bar z,t)
\right] = \chi^\prime(t) = -{\bar\chi}^\prime(t) , \]
where $\chi^\prime(t)$ is the separation `constant" and the prime denotes
the derivative.
 Solving these equations with respect to $\psi,\bar\psi$
and substituting the results into the equations (\ref{intYeq}) we solve them
with respect to the square root with the result
\[\sqrt{2\kappa\rho-u_t^2}=i\kappa\left[\alpha_t(z,t)-\bar\alpha_t(\bar z,t)
-2\chi^\prime(t)\right]. \]
Solving this equation with respect to $\kappa\rho$ and multiplying the result
by $e^{\alpha+\bar\alpha}$ we obtain
\begin{eqnarray}
\!\!&\!\!\!\!&\!\! \kappa e^{-\gamma(z,\bar z)}\gamma_{z\bar z}(z,\bar z)
\label{roeq}
\\ \!\!&\!\!\!\!&\!\!  = 2e^{\alpha(z,t)+\bar\alpha(\bar z,t)}
\left[\alpha_t(z,t) \bar\alpha_t(\bar z,t)
+\chi^\prime(t)\bigl(\alpha_t(z,t) -\bar\alpha_t(\bar z,t)\bigr)
-{\chi^\prime}^2(t) \right].
\nonumber
\end{eqnarray}
Using this equation in the right-hand side of the heavenly equation in the
form (\ref{heavmod}) and separating $z,\bar z$ we obtain two complex
conjugate equations
\begin{eqnarray}
\!\!&\!\!\!\!&\!\! \alpha_{tt}(z,t) = -\alpha_t^2(z,t)
+ 2\chi^\prime(t)\alpha_t(z,t) - {\chi^\prime}^2(t) + \mu(t) ,
\label{separ}
\\ \!\!&\!\!\!\!&\!\! \bar\alpha_{tt}(\bar z,t) =
- {\bar\alpha}_t^2(\bar z,t) - 2\chi^\prime(t)\bar\alpha_t(\bar z,t)
- {\chi^\prime}^2(t) - \mu(t) ,
\label{bsepar}
\end{eqnarray}
where $\mu(t)=-\bar\mu(t)$ is the separation ``constant". We substitute
these expressions for $\alpha_{tt}$ and $\bar\alpha_{tt}$ into the
transformed heavenly equation (\ref{heavmod}) to obtain
\begin{eqnarray}
\!\!&\!\!\!\!&\!\!
e^{\alpha(z,t)+\bar\alpha(\bar z,t)}\left[\alpha_t(z,t)\bar\alpha_t(\bar z,t)
+\chi^\prime(t)\bigl(\alpha_t(z,t)-\bar\alpha_t(\bar z,t)
-{\chi^\prime}^2(t)\bigr)\right]
\nonumber
\\ \!\!&\!\!\!\!&\!\! = \frac{\kappa}{2}\,e^{-\gamma(z,\bar z)}
\gamma_{z\bar z}(z,\bar z).
\label{heavsubst}
\end{eqnarray}
Next we take the total derivative $D_t$ of this equation and substitute again
the second derivatives $\alpha_{tt}$ and $\bar\alpha_{tt}$ from the
equations (\ref{separ}) and (\ref{bsepar}). The result is unexpectedly
simple
\begin{equation}
\left(\chi^{\prime\prime}-\mu\right)\left(\alpha_t-\bar\alpha_t
-2\chi^\prime\right) = 0.
\label{simpl}
\end{equation}
This equation implies that
\begin{equation}
\mu(t) = \chi^{\prime\prime}(t) ,
\label{mueq}
\end{equation}
since the complementary assumption
\[ \alpha_t-\bar\alpha_t -2\chi^\prime = 0 \]
leads again to the equation (\ref{mueq}). Indeed, the last equation allows
a separation of $z,\bar z$ and, being integrated, gives $\alpha,\bar\alpha$
\[ \alpha(z,t) = \chi(t)+\nu(t)+\omega(z), \qquad
\bar\alpha(\bar z,t) = -\chi(t)+\nu(t)+\bar\omega(\bar z).\]
Substituting these expressions into the equations (\ref{separ}) and
(\ref{bsepar}) and comparing the results we discover again the equation
(\ref{mueq}).

  With this restriction the equations (\ref{separ}) and (\ref{bsepar}) are
simplified and integrated to give
\begin{equation}
\alpha(z,t) = \ln{\bigl(t+b(z)\bigr)} + \chi(t) + \omega(z), \quad
\bar\alpha(\bar z,t) = \ln{\bigl(t+\bar a(\bar z)\bigr)} - \chi(t) +
\bar\omega(\bar z),
\label{alpsol}
\end{equation}
where $b(z)$ and $\omega(z)$ are arbitrary holomorphic functions
and we have reserved the notation $a(z)$ only for the
generators of the conformal vector field $X_a$ in eq.(\ref{symgen}).

Now define a new function of $z,\bar z$
\begin{equation}
\Gamma(z,\bar z) = \gamma(z,\bar z) + \omega(z) + \bar\omega(\bar z),
\label{defGam}
\end{equation}
so that the form (\ref{solform}) of the solution becomes
\begin{equation}
u(z,\bar z,t) = \ln{\bigl(t+b(z)\bigr)} + \ln{\bigl(t+\bar b(\bar z)\bigr)}
+ \Gamma(z,\bar z).
\label{solformnew}
\end{equation}
Substituting the expressions (\ref{alpsol}) for $\alpha,\bar\alpha$ into
the transformed heavenly equation (\ref{heavmod}) we obtain the equation
for the only unknown function $\Gamma(z,\bar z)$ in the solution
(\ref{solformnew})
\begin{equation}
\Gamma_{z\bar z} = 2\kappa e^\Gamma.
\label{Liouville}
\end{equation}
If $\kappa=1$ this is the Liouville equation with the general solution
\begin{equation}
\Gamma(z,\bar z) = \ln{c^\prime(z)} + \ln{{\bar c}^\prime(\bar z)}
- 2 \ln{\bigl(c(z)+\bar c(\bar z)\bigr)}
\label{Gplus}
\end{equation}
where $c(z)$ is an arbitrary holomorphic function.
If $\kappa=-1$ we call the equation (\ref{Liouville})
\`{}$\!$pseudo-Liouville$\!$\'{} equation and its general solution is
\begin{equation}
\Gamma(z,\bar z) = \ln{c^\prime(z)} + \ln{{\bar c}^\prime(\bar z)}
- 2 \ln{\bigl(c(z)\bar c(\bar z)+1\bigr)}.
\label{Gminus}
\end{equation}
Finally, substituting these expressions for $\Gamma(z,\bar z)$ into the
equation (\ref{solformnew}) we obtain the solutions of the heavenly equation
(\ref{heav}) for the two choices of the sign $\kappa=+1$ and $\kappa=-1$.
\begin{enumerate}
\item
The solution for $\kappa=1$:
\begin{eqnarray}
\!\!&\!\!\!\!&\!\! u(z,\bar z,t) =
\ln{\bigl(t+b(z)\bigr)}+\ln{\bigl(t+\bar b(\bar z)\bigr)}
\nonumber
\\ \!\!&\!\!\!\!&\!\! \mbox{} +\ln{c^\prime(z)} +\ln{{\bar c}^\prime(\bar z)}
- 2 \ln{\bigl(c(z)+\bar c(\bar z)\bigr)}.
\label{solplus}
\end{eqnarray}
\item
The solution for $\kappa=-1$ (see also \cite{tod}):
\begin{eqnarray}
\!\!&\!\!\!\!&\!\! u(z,\bar z,t) =
\ln{\bigl(t+b(z)\bigr)}+\ln{\bigl(t+\bar b(\bar z)\bigr)}
\nonumber
\\ \!\!&\!\!\!\!&\!\! \mbox{}+\ln{c^\prime(z)} + \ln{{\bar c}^\prime(\bar z)}
- 2 \ln{\bigl(c(z)\bar c(\bar z)+1\bigr)}.
\label{solminus}
\end{eqnarray}
\end{enumerate}
Here $b(z)$ and $c(z)$ are arbitrary holomorphic functions.

  To avoid ``false generality" it is sufficient to choose the simplest
representative of the obtained orbits of solutions applying the conformal
symmetry transformation $c(z)=z,\;\bar c(\bar z)=\bar z$ with the following
results.
\begin{enumerate}
\item
The solution for $\kappa=1$:
\begin{equation}
u(z,\bar z,t) = \ln{\bigl(t+b(z)\bigr)}+\ln{\bigl(t+\bar b(\bar z)\bigr)}
 - 2 \ln{(z+\bar z)}.
\label{simpsolpl}
\end{equation}
\item
The solution for $\kappa=-1$:
\begin{equation}
u(z,\bar z,t) = \ln{\bigl(t+b(z)\bigr)}+\ln{\bigl(t+\bar b(\bar z)\bigr)}
 - 2 \ln{(z\bar z+1)}.
\label{simpsolmin}
\end{equation}
\end{enumerate}
Here $b(z)$ is still an arbitrary holomorphic function.

  Up to now we solved completely only the Ansatz (\ref{ans})
defining $\tau$, but we did not check the automorphic equation (\ref{autom})
and the auxiliary equations (\ref{lambdef}), by using
the particular solution (\ref{solvresolv}) of the
resolving system. Hence, though we obtained the correct
solutions (\ref{simpsolpl}) and (\ref{simpsolmin}) of the heavenly equation
(\ref{heav}), we have not made a complete foliation of these solutions into
separate orbits.

  To do this, first we remark that due to the discrete symmetry of the
heavenly equation (\ref{heav}) and of our solutions with respect to the
permutation $z\leftrightarrow \bar z$, we can define the
holomorphic function $b(z)$ as satisfying the condition
\begin{equation}
{\rm Im}\,b(z)\ge 0\;{\rm for}\;\kappa=1 \quad {\rm and} \quad
{\rm Im}\,b(z)\le 0\;{\rm for}\;\kappa=-1.
\label{condit}
\end{equation}
Then we check that the automorphic equation (\ref{autom}) coincides with the
auxiliary equations (\ref{lambdef}) and becomes
\begin{equation}
(z+\bar z)^2 b^\prime(z) {\bar b}^\prime(\bar z)=8\varphi(\xi,\theta)\qquad
{\rm for}\quad \kappa=1
\label{autpl}
\end{equation}
and
\begin{equation}
(z\bar z+1)^2 b^\prime(z) {\bar b}^\prime(\bar z)=-8\varphi(\xi,\theta)\qquad
{\rm for}\quad\kappa=-1 .
\label{autmin}
\end{equation}
Using the solutions (\ref{simpsolpl}) and (\ref{simpsolmin}) in the
definitions (\ref{sol1}) of the characteristic variables $\xi$ and $\theta$,
we discover that they depend only on $b$ and $\bar b$, {\it i.e.}
\begin{equation}
\xi = -\frac{(b-\bar b)^2}{4},\qquad\theta = -\frac{(b+\bar b)}{2} -\sqrt{\xi}.
\label{xieta}
\end{equation}
Hence, defining the new arbitrary function
$\Phi(b,\bar b)=\varphi(\xi,\theta)$, the automorphic equations (\ref{autpl})
and (\ref{autmin}) become
\begin{equation}
(z+\bar z)^2 b^\prime(z) {\bar b}^\prime(\bar z) = 8\Phi(b,\bar b)\qquad
{\rm for}\quad \kappa=1
\label{aut+}
\end{equation}
and
\begin{equation}
(z\bar z+1)^2 b^\prime(z) {\bar b}^\prime(\bar z) = -8\Phi(b,\bar b)\qquad
{\rm for}\quad \kappa=-1.
\label{aut-}
\end{equation}
Sufficient conditions for solving these functional-differential equations
are given by the following choices of $\Phi(b,\bar b)$
\begin{equation}
\Phi(b,\bar b)=
\frac{\left[f(b)+\bar f(\bar b)\right]^2}{8f^\prime(b){\bar f}^\prime(\bar b)}
\qquad {\rm for}\quad \kappa=1
\label{phi+}
\end{equation}
and
\begin{equation}
\Phi(b,\bar b)=-\frac{\left[f(b)\bar f(\bar b)
+1\right]^2}{8f^\prime(b){\bar f}^\prime(\bar b)}\qquad {\rm for}\quad
\kappa=-1,
\label{phi-}
\end{equation}
where $f(b)$ is an arbitrary holomorphic function.
Then the automorphic equations become
\begin{equation}
\left\{\ln{\left[f(b)+\bar f(\bar b)\right]}\right\}_{z\bar z}
=\left[\ln{(z+\bar z)}\right]_{z\bar z}\qquad {\rm for}\quad \kappa=1
\label{autom+}
\end{equation}
and
\begin{equation}
\left\{\ln{\left[f(b)\bar f(\bar b)+1\right]}\right\}_{z\bar z}
=\left[\ln{(z\bar z+1)}\right]_{z\bar z}\qquad {\rm for}\quad \kappa=-1 .
\label{autom-}
\end{equation}
Their general solutions are
\begin{equation}
f(b)+\bar f(\bar b) = w(z)\bar w(\bar z)(z+\bar z)
\qquad {\rm for}\quad \kappa=1
\label{autsol+}
\end{equation}
and
\begin{equation}
f(b)\bar f(\bar b)+1 = w(z)\bar w(\bar z)(z\bar z+1)\qquad {\rm for}\quad
\kappa=-1 ,
\label{autsol-}
\end{equation}
where $w(z)$ is an arbitrary holomorphic function.
The formulae (\ref{phi+}) and (\ref{phi-}) for $\Phi(b,\bar b)$ become
\[\Phi(b,\bar b) = \frac{w^2(z){\bar w}^2(\bar z)(z+\bar
z)^2}{8f^\prime(b){\bar f}^\prime(\bar b)} \qquad {\rm for}\quad \kappa=1\]
and
\[\Phi(b,\bar b) = -\frac{w^2(z){\bar w}^2(\bar z)(z\bar
z+1)^2}{8f^\prime(b){\bar f}^\prime(\bar b)} \qquad {\rm for}\quad
\kappa=-1.\]
If we plug these formulae into the automorphic equations (\ref{aut+}) and
(\ref{aut-}), then both automorphic equations coincide and become
\begin{equation}
b^\prime(z){\bar b}^\prime(\bar z)=
\frac{w^2(z){\bar w}^2(\bar z)}{f^\prime(b){\bar f}^\prime(\bar b)}\,.
\label{automorph}
\end{equation}
This equation admits a separation of variables, leading to the ODEs
\begin{equation}
b^\prime(z)=\frac{w^2(z)}{f^\prime(b)},\qquad
{\bar b}^\prime(z)=\frac{{\bar w}^2(\bar z)}{{\bar f}^\prime(\bar b)}\,.
\label{autODE}
\end{equation}
The obvious choice of the functions $w(z)$ and $\bar w(\bar z)$
\begin{equation}
w(z)=1 \quad\Longleftrightarrow \quad\bar w(\bar z)=1
\label{wsol}
\end{equation}
simplifies the ODEs (\ref{autODE}) to
\begin{equation}
[f(b)]_z=1, \qquad [\bar f(\bar b)]_{\bar z}=1 ,
\label{simpODE}
\end{equation}
with the solution
\begin{equation}
f[b(z)]=z, \qquad \bar f[\bar b(\bar z)]=\bar z
\label{solODE}
\end{equation}
meaning that $b(z)$ is the inverse function for $f(b)$: $b=f^{-1}$. The
equations (\ref{autsol+}) and (\ref{autsol-}) are obviously satisfied by
the solution (\ref{solODE}) with our choice (\ref{wsol}) of $w(z),\bar
w(\bar z)$.

  Thus, any particular function $b(z)$ can be obtained for an appropriate
choice of $f(b)$ as its inverse function. This fixes the function
$\Phi(b,\bar b)$ according to the formulae (\ref{phi+}) or (\ref{phi-}),
the function $\varphi(\xi,\theta)=\Phi(b,\bar b)$ and the right-hand side $F$
of the automorphic equation (\ref{autom}) determined by the formulae
(\ref{solvresolv}). Hence any particular choice of the function $b(z)$
in our solutions (\ref{simpsolpl}), (\ref{simpsolmin}) means
a corresponding choice of the particular orbit in the solution space of the
heavenly equation.

\section{Check of non-invariance of the solutions}
\label{sec:check}

\setcounter{equation}{0}

 In this section we prove that our solutions (\ref{simpsolpl}) and
(\ref{simpsolmin}) of the heavenly equation (\ref{heav}) are non-invariant,
with respect to its symmetry group generated by the vector fields in
(\ref{symgen}), for
generic functions $a(z)$, except for some particular classes
listed below in the theorems summarizing the results.

 For the check of non-invariance we substitute our solutions (\ref{simpsolpl})
and (\ref{simpsolmin}) into the invariance criterion (\ref{critinv}).
The resulting equation is quadratic in $t$ and it implies the vanishing of
the coefficients of $t^2,t$ and $t^0$.

  We consider first the case $\kappa=1$. The term with $t^2$ gives
again the equation (\ref{odeplus}) of the Section 6.
However, now we need the general solution of this equation.

  We assume in the generic case that $a^\prime(z)+{\bar a}^\prime(\bar z)\ne
0$, otherwise the equation (\ref{odeplus}) implies $a=-\bar a=constant$ and
this case should be treated separately. Then we rewrite the equation
(\ref{odeplus}) in the form
\begin{equation}
\frac{a(z)+\bar a(\bar z)}{a^\prime(z)+{\bar a}^\prime(\bar z)} =
\frac{z+\bar z}{2} \qquad\Longrightarrow\qquad
\left(\frac{a+\bar a}{a^\prime+{\bar a}^\prime}\right)_{z\bar z} =0 .
\label{modt^2+}
\end{equation}
In order to consider the generic case we postulate
$a^{\prime\prime}{\bar a}^{\prime\prime}\ne0$, then the last equation can be
easily manipulated, obtaining the solution
\begin{equation}
a(z) = C_1(z+\lambda)^2+C_2,\qquad \bar a(\bar z) =
-\left[C_1(\bar z-\lambda)^2+C_2\right] ,
\label{a,ba}
\end{equation}
where $C_1\ne 0$, $C_2$ and $\lambda$ are arbitrary purely imaginary
constants.

  Now we consider the term without $t$ in the criterion of invariance using
our result (\ref{a,ba}) which gives the equation with the separated variables
$z,\bar z$
\begin{eqnarray}
\!\!&\!\!\!\!&\!\!\frac{\alpha}{b(z)}+\left[C_1(z+\lambda)^2+C_2\right]
\frac{b^\prime(z)}{b(z)}-\beta =
\label{t^0+}
\\ \!\!&\!\!\!\!&\!\! -\left\{
\frac{\alpha}{\bar b(\bar z)}-\left[C_1(\bar z-\lambda)^2+C_2\right]
\frac{{\bar b}^\prime(\bar z)}{\bar b(\bar z)}-\beta\right\}
=\mu=-\bar\mu  \nonumber
\end{eqnarray}
where $\mu$ is a separation constant.
Comparing these equations with the equation obtained from the term with $t$
in the criterion of invariance we conclude that they coincide if and only if
the condition
\[ \mu\left(b(z)-\bar b(\bar z)\right) = 0 \]
is satisfied. This implies $\mu=0$, since otherwise we have
$b=\bar b=constant$ and our solution is obviously invariant depending only on
two variables $t$ and $z+\bar z$. Hence the equations (\ref{t^0+}) become
\begin{eqnarray}
\!\!&\!\!\!\!&\!\!\left[C_1(z+\lambda)^2+C_2\right]b^\prime(z)-\beta b(z)
=-\alpha ,
\label{t+}
\\[1mm]
\!\!&\!\!\!\!&\!\!\left[C_1(\bar z-\lambda)^2+C_2\right]
{\bar b}^\prime(\bar z) + \beta \bar b(\bar z)=\alpha .
\label{bt+}
\end{eqnarray}
Consider now the case $C_2\ne 0,\;\beta\ne 0$ and introduce the notation
\begin{equation}
\nu=\sqrt{-\frac{C_2}{C_1}},\qquad \gamma=\frac{\beta}{2\sqrt{-C_1C_2}}\,.
\label{notation+}
\end{equation}
Integrating the ODEs (\ref{t+}) and (\ref{bt+}) we fix the function $b(z)$
in our solution of the heavenly equation which corresponds to the invariant
solution in the considered case
\begin{equation}
b(z) = C\left(\frac{z+\lambda-\nu}{z+\lambda+\nu}\right)^\gamma +
\frac{\alpha}{\beta},\quad
\bar b(\bar z) = \bar C\left(\frac{\bar z-\lambda-\nu}{\bar z-\lambda+\nu}
\right)^\gamma + \frac{\alpha}{\beta}
\label{b+}
\end{equation}
where $C,\bar C$ are integration constants.

 In a similar way we treat other possible cases. We sum up the results for
the case of $\kappa=1$ in the following statement.
\begin{theorem}
The function
\begin{equation}
u = \ln{\bigl(t+b(z)\bigr)}+\ln{\bigl(t+\bar b(\bar z)\bigr)}
 - 2 \ln{(z+\bar z)}
\label{sol+}
\end{equation}
is a solution of the heavenly equation (\ref{heav}) for $\kappa=+1$
for an arbitrary holomorphic function $b(z)$. This solution is a
non-invariant solution of this equation iff the function $b(z)$ does not
coincide with any of the following choices:
\begin{enumerate}
\item
\[b(z) = C\left(\frac{z+\lambda-\nu}{z+\lambda+\nu}\right)^\gamma +
\frac{\alpha}{\beta}\]
where $\alpha$ and $\beta$ are arbitrary real constants, $\beta\ne 0$,
$\nu$ and $\gamma$ are defined by the formulae (\ref{notation+}) and
$\lambda,C_1,C_2$ are complex constants which satisfy the conditions
\[ \bar\lambda=-\lambda,\quad \bar C_1=-C_1,\quad \bar C_2=-C_2,
\quad C_1\ne 0,\quad C_2\ne 0. \]
In this case the solution is invariant with respect to the symmetry generator
\begin{eqnarray*}
\!\!&\!\!\!\!&\!\!X = \alpha\partial_t +\beta\left(t\partial_t+2\partial_u
\right)+C_1\left[(z+\lambda)^2\partial_z-(\bar z-\lambda)^2\partial_{\bar z}
- 2(z-\bar z)\partial_u\right]
\\ \!\!&\!\!\!\!&\!\!\mbox{}+C_2 \left(\partial_z -\partial_{\bar z}\right).
\end{eqnarray*}
\item
\[b(z) =
\frac{\alpha}{2\sqrt{-C_1C_2}}
\left(\frac{z+\lambda+\nu}{z+\lambda-\nu}\right) + C\qquad {\rm if}\quad
\beta=0,\;C_2\ne 0;\]
the solution is invariant with respect to the previous symmetry generator $X$
with $\beta=0$.
\item
\[ b(z)=C\exp{\left[-\frac{\beta}{C_1(z+\lambda)}\right]}
+\frac{\alpha}{\beta}\qquad {\rm if}\quad C_2=0,\;\beta\ne 0; \]
the solution is invariant with respect to the symmetry generator $X$ from the
case $1$ with $C_2=0$.
\item
\[ b(z)=\frac{\alpha}{C_1(z+\lambda)} + C\qquad {\rm if}\quad \beta=0\;
{\rm and}\; C_2= 0;\]
the solution is invariant with respect to the symmetry generator $X$ from the
case $1$ with $\beta=0$ and $C_2=0$.
\item
\[ b(z) = C(C_1z+C_2)^{\beta/C_1}+\frac{\alpha}{\beta}\qquad {\rm if}\quad
C_1\ne 0,\;\beta\ne 0; \]
the solution is invariant with respect to the symmetry generator
\[ X = \alpha\partial_t +\beta\left(t\partial_t+2\partial_u\right)
+C_1\left(z\partial_z+\bar z\partial_{\bar z}-2\partial_u\right)
+C_2 \left(\partial_z -\partial_{\bar z}\right).\]
\item
\[ b(z) = Ce^{\frac{\beta}{C_2}z}+\frac{\alpha}{\beta}
\qquad {\rm if}\quad C_1 = 0,\;\beta\ne 0;\]
the solution is invariant with respect to the symmetry generator $X$ from the
case $5$ with $C_1=0$.
\item
\[ b(z) = -\frac{\alpha}{C_2}\,z+C\qquad {\rm if}\quad C_1 = 0,\;\beta = 0;\]
the solution is invariant with respect to the symmetry generator $X$ from the
case $5$ with $C_1=0$ and $\beta=0$.
\item
\[ b(z) = b = constant \qquad {\rm if}\quad C_1=\alpha=\beta=0,\;C_2\ne 0;\]
the solution is invariant with respect to the symmetry generator
\[ X=\partial_z - \partial_{\bar z}. \]
If $b=\alpha/\beta$, then this solution is also invariant with respect to the
generator
\[ X = \alpha\partial_t +\beta\left(t\partial_t+2\partial_u\right). \]
\end{enumerate}
\end{theorem}

  Now we consider the case $\kappa=-1$ and substitute the solution
(\ref{simpsolmin}) of the heavenly equation (\ref{heav})
into the criterion of invariance (\ref{critinv}). Then
the resulting equation is again quadratic in $t$ and the term with $t^2$
gives us again the equation (\ref{odemin}),
for which we need now the general solution.
First we rewrite it in the form
\[ a^\prime+{\bar a}^\prime + \frac{a^\prime + {\bar a}^\prime}{z\bar z}
= 2\left(\frac{a}{z}+\frac{\bar a}{\bar z}\right). \]
Differentiating this equation with respect to $z$ and $\bar z$ we obtain
an equation which admits separation of $z,\bar z$ in the form
\begin{equation}
za^{\prime\prime}(z)-a^\prime(z)=-\left[\bar z{\bar a}^{\prime\prime}(\bar z)
-{\bar a}^\prime(\bar z)\right]=\lambda=-\bar\lambda
\label{ODEs}
\end{equation}
where $\lambda$ is a separation constant. Integrating these ODEs we obtain
\[ a(z)=C_1z^2-\lambda z+C_2,\qquad \bar a(\bar z)=\bar C_1{\bar z}^2
+\lambda\bar z+\bar C_2 \]
where $C_1,C_2$ are integration constants. Substituting these solutions into
the equation (\ref{odemin}) we see that it is identically satisfied if and only
if \newline $\bar C_1=C_2\;\iff \;\bar C_2=C_1$, so that finally we have the
solution of the equation following from the term with $t^2$
\begin{equation}
a(z) = C_1z^2-\lambda z+C_2,\qquad \bar a(\bar z)=C_2{\bar z}^2+\lambda\bar z
+C_1.
\label{a,ba-}
\end{equation}

  Next we consider the term without $t$ in the criterion of invariance using
our result (\ref{a,ba-}) which gives the equation with the separated variables
$z,\bar z$
\begin{eqnarray}
\!\!&\!\!\!\!&\!\!\frac{\alpha}{b(z)}+\left(C_1z^2-\lambda z+C_2\right)
\frac{b^\prime(z)}{b(z)}-\beta =  \nonumber
\\ \!\!&\!\!\!\!&\!\! -\left[
\frac{\alpha}{\bar b(\bar z)}+\left(C_2{\bar z}^2+\lambda\bar z+C_1\right)
\frac{{\bar b}^\prime(\bar z)}{\bar b(\bar z)}-\beta\right]=\mu=-\bar\mu
\label{t^0-}
\end{eqnarray}
where $\mu$ is a separation constant.
Comparing these equations with the equation obtained from the term with $t$
in the criterion of invariance we conclude that they coincide if and only if
the condition
\[ \mu\left(b(z)-\bar b(\bar z)\right) = 0 \]
is satisfied. This implies $\mu=0$ for the same reason as in the case
$\kappa=1$.  Hence the equations (\ref{t^0-}) become
\begin{eqnarray}
\!\!&\!\!\!\!&\!\!\left(C_1z^2-\lambda z+C_2\right)b^\prime(z)-\beta b(z)
=-\alpha ,
\label{t-}
\\[1mm]
\!\!&\!\!\!\!&\!\!\left(C_2{\bar z}^2+\lambda\bar z+C_1\right)
{\bar b}^\prime(\bar z) - \beta \bar b(\bar z)=-\alpha .
\label{bt-}
\end{eqnarray}
Consider now the case $C_1\ne 0$. Introduce the new constants
$\tilde\lambda=-\lambda/(2C_1)$ and $\tilde C_2=C_2-\lambda^2/(4C_1)$.
Then the first equation takes the form
\begin{equation}
\left[C_1(z+\tilde\lambda)^2+\tilde C_2\right]b^\prime(z)-\beta b(z)=-\alpha
\label{tt-}
\end{equation}
coinciding with the ODE (\ref{t+}) in the case $\kappa=1$. Therefore we can
use its solutions with an appropriate change of notation. Other possible
cases are treated in a similar way. Therefore we can transfer the results
of Theorem \thetheorem\ to the case $\kappa=-1$ with an appropriate
change of notation and sum them up in the following statement.
\begin{theorem}\addtocounter{theorem}{-1}
The function
\begin{equation}
u = \ln{\bigl(t+b(z)\bigr)}+\ln{\bigl(t+\bar b(\bar z)\bigr)}
 - 2 \ln{(z\bar z+1)}
\label{sol-}
\end{equation}
is a solution of the heavenly equation (\ref{heav}) for $\kappa=-1$
for an arbitrary holomorphic function $b(z)$. This solution is a
non-invariant solution of this equation iff the function $b(z)$ does not
coincide with any of the $8$ forms given in Theorem \thetheorem\ with
the change of notation
\begin{eqnarray*}
\!\!&\!\!\!\!&\!\!\lambda\mapsto \tilde\lambda=-\frac{\lambda}{2C_1},\quad
C_2\mapsto \tilde C_2=C_2-\frac{\lambda^2}{4C_1},
\\ \!\!&\!\!\!\!&\!\!\nu\mapsto \tilde\nu =
\sqrt{-\frac{\tilde C_2}{C_1}},\quad\gamma\mapsto \tilde\gamma=
\frac{\beta}{2\sqrt{-C_1\tilde C_2}}
\end{eqnarray*}
in the cases $1,2,3,4$ and $C_1\mapsto -\lambda$ in the case $5$. Those $8$
choices of $b(z)$ give invariant solutions with respect to the corresponding
symmetry generators of Theorem \thetheorem\ with the same change of
notation.
\end{theorem}
\addtocounter{theorem}{1}

\section{Conclusions and outlook}
\label{sec:conclusion}
\setcounter{equation}{0}

The title of this article, or rather of the research direction that it
represents, could have been ``Invariant methods for obtaining non-invariant
solutions of partial differential equations". The main result is that we are
proposing an alternative tool for obtaining particular solutions of
non-linear partial differential equations with infinite dimensional symmetry
algebras. As stated in the Introduction, the idea of the method is more than
a hundred years old \cite{lie,vessiot}. We have turned it into a usable tool
by adding new elements. These are:
\begin{enumerate}
\item
The systematic use of invariant cross-differentiation involving
the operators of invariant differentiation and their commutator
algebra for the derivation of the resolving equations and for
obtaining their particular solutions.
\item
The presentation of the resolving system as a Lie algebra of the
operators of invariant differentiation (over the field of differential
invariants of the symmetry group) \cite{ns}.
\item
The concept of invariant integration applied to the automorphic system.
\end{enumerate}

  Let us use the heavenly equation (\ref{heav}) to compare different
methods of obtaining exact analytical solutions of a partial differential
equation, provided or at least suggested by symmetry analysis. In all of them
the studied equation is embedded into a larger system of equations, to be
solved simultaneously.

  The most standard method is that of invariant solutions \cite{ovs,olv,wint}.
One first finds the symmetry algebra realized by vector fields of the form
\begin{equation}
X = \tau\partial_t +\xi\partial_z +\bar\xi\partial_{\bar z} +\phi\partial_u
\label{vfield}
\end{equation}
where $\tau,\xi,\bar\xi$ and $\phi$ are functions of $t,z,\bar z$ and $u$.
Once this algebra is found ({\it i.e.} the algebra (\ref{symgen}) for the
heavenly equation) one classifies its subalgebras into conjugacy classes and
then adds one, or more, first order linear equations of the type
\begin{equation}
\tau u_t +\xi u_z + \bar\xi u_{\bar z} -\phi = 0
\label{addeq}
\end{equation}
to the studied equation. These equations are solved, their solution is
substituted into the original equation. This again is solved and we obtain
solutions invariant under the chosen subgroup.

  Further methods are the Bluman and Cole ``non-classical method"
\cite{blucol}, the Clarkson-Kruskal \cite{clarkrus} ``direct method" and that
of ``conditional symmetries" \cite{levwin} (see \cite{clarwint} for a review).
These methods, basically all equivalent, amount to the fact that a first
order equation of the type (\ref{addeq}) is added to the studied equation,
without the requirement that $\tau,\xi,\bar\xi$ and $\phi$ define an element
of the symmetry algebra.

  Finally, we have the group foliation method \cite{ns} used and further
developed in this article. Let us review the essential steps, performed
above.
\begin{enumerate}
\item
Find the total symmetry algebra (\ref{symgen}).
\item
Find all differential invariants of order up to $N$ of its
infinite dimensional subalgebra which is Lie algebra of the
conformal group. The number $N$ must be larger or equal to the
order of the equation and must satisfy the requirement that there
should be $\# N$ functionally independent invariants with
\begin{equation}
\# N \ge p+q
\label{ineq}
\end{equation}
where $p$ and $q$ are the number of independent and dependent variables,
respectively. In our case we have $p=3,\; q=1,\; N=2,\; \# N=5$. The actual
invariants are given in the equation (\ref{difinv}).
\item
Choose $p$ invariants as new independent variables and require
that the remaining invariants be functions of the chosen ones.
This provides us with the automorphic system that also contains
the considered equation, expressed in terms of the invariants. In
our case the automorphic system consists of the equation
(\ref{heavinv}) (the heavenly equation) and the equation
(\ref{autom}) (or equivalently (\ref{Yautom})).
\item
Find the ``resolving equations". This is a set of compatibility conditions
between the studied equation and those that we have added to obtain the
automorphic system. In our case we require compatibility between the
equations (\ref{heavinv}) and (\ref{autom}), {\it i.e.} determine the
restrictions on the function $F(t,u_t,\rho)$. We have shown that this can be
done in an explicitly invariant manner by using the operators of invariant
differentiation, in our case $\delta,Y$  and $\bar Y$ of the equations
(\ref{deltas}) and (\ref{Ydef}). The resolving system in our case consists of
the equations (\ref{1}), (\ref{3}), (\ref{2}), (\ref{b2}) and (\ref{4}). As
stated by the fundamental Theorem $1$, this resolving system is best written
as a system of commutator relations for the operators of invariant
differentiation projected on the solution manifold of the heavenly equation
in the space of differential invariants, together with the Jacobi relations
for these operators.
\item
Solve the resolving system and the automorphic one. This provides solutions
of the original equation.
\end{enumerate}

  The last step, step $5$ is the most difficult one. If it can be carried out
completely, we obtain ``all" solutions, both invariant and
non-invariant ones. In general, such a situation is too good to be
true. In particular, for the heavenly equation we were not able to
solve the system (\ref{Yalg}), (\ref{jacobi}) in general. Instead,
we made various simplifying assumptions. The most obvious ones,
like $Y=\bar Y$ or $F=0$, lead to invariant solutions. These we
already know, or can obtain by much simpler standard methods. The
assumption, or restriction, that leads to non-invariant solutions
was $[Y,\bar Y]=0$. The solutions obtained are (\ref{solplus}) and
(\ref{solminus}), for $\kappa=1$ and $\kappa=-1$, respectively.
Each solution involves two arbitrary holomorphic functions. One of
them, $b(z)$ is fundamental. The other is induced by a
conformal transformation and can be transformed away ({\it i.e.}
set equal to {\it e.g.} $c(z)=z$). In Section \ref{sec:check} we
show that the solutions are, in general, not invariant under any
subgroup of the symmetry group. They reduce to invariant ones only
for very special choices of the function $b(z)$, specified in
Theorems $2$ and $3$.

  It would be interesting to relate the concepts of this article to that of
integrability for non-linear partial differential equations. ``Integrability"
means that the considered equation is viewed as an integrability condition
for a Lax pair, a pair of linear operators \cite{lax,ablclar}. Here we can
view the equations (\ref{comalg}) as a set of relations between a triplet of
linear operators, subject to a non-linear constraint (\ref{jacobi}) .

\section*{Acknowledgments}

  A large part of the research reported here was performed while M.B.S. and
P.W. were visiting the Universit\`{a} di Lecce. They thank the
Dipartimento di Fisica and INFN, Sezione di Lecce, for their
hospitality and support. One of the authors (M.B.S.) thanks Y.
Nutku for useful discussions.

The research of P.W. is partly supported by research grants from
NSERC of Canada and FCAR du Qu\'{e}bec. The research of L.M. is
supported by INFN - Sezione di Lecce and by the research grant
Prin - Sintesi 2000 from MURST of Italy and it is a part of the
INTAS research project $99 - 1782$.

\end{document}